# PHOTOLUMINESCENCE TUNING THROUGH IRRADIATION DEFECTS IN $CH_3NH_3PBI_3$ PEROVSKITES

Olivier Plantevin, Stéphanie Valère, Driffa Guerfa, Ferdinand Lédée, Gaëlle Trippé-Allard, Damien Garrot, Emmanuelle Deleporte

Dr. O. Plantevin, S. Valère, D. Guerfa

Centre de Sciences Nucléaires et de Sciences de la Matière, CNRS-IN2P3-Univ. Paris-Sud, Université Paris-Saclay, 91405 Orsay Cedex, France

E-mail : plantevin@csnsm.in2p3.fr

Dr. D. Garrot, Groupe d'Etude de la Matière Condensée, CNRS-Université de Versailles St-Quentin en Yvelines, Université Paris-Saclay, 45 av. des Etats-Unis, 78035 Versailles, France

Dr. F. Lédée, Dr. G. Trippé-Allard, Prof. E. Deleporte, Laboratoire Aimé Cotton, CNRS-Univ Paris-Sud, ENS Paris-Saclay, Université Paris - Saclay, 91405 Orsay Cedex, France

Abstract

Defect engineering is applied to hybrid $(CH_3NH_3)PbI_3$ organic-inorganic perovskites. These materials have become one of the most promising low-cost alternatives to traditional semiconductors in the field of photovoltaics and light emitting devices. We use here Helium ion irradiation at low energy as a tool for the controlled introduction of point defects in both single crystals and polycrystalline thin films. The irradiation defects modify the opto-electronic properties as probed using photoluminescence spectroscopy from 10K to room-temperature.

Contrary to usual semiconductors,  we observe a very good resilience of the PL properties with irradiation, even associated to an enhancement of the optical emission at low temperature. We discuss these results in relation with the tetragonal to orthorhombic low-temperature phase transition below T=160 K. A comparison between spectra from single crystals and polycrystalline films, both with and without irradiation defects, allow a better understanding of the light emission mechanisms in both kinds of samples. We thereby evidence radiation hardness of these materials and the specificity of defects and their impact over light emission properties.

Introduction

In materials science, as seen in particular in semiconductor physics, it is crucial to work on the relationship between the structural, electronic and optical properties. We are considering here more specifically issues in hybrid perovskite materials. Can we understand better these semiconductor materials' properties like light emission with the help of defects engineering?[1] Recently, defect tolerance and defect passivation have become a challenge to improve efficiencies in the frame of photovoltaic devices.[2,3] In the past two decades, the family of hybrid organic-inorganic perovskites (HOP) have arisen as new functional materials firstly for their light emitting properties, with the structure $A_2MX_4$ (2D perovskites) where A is an organic cation, M is a metal cation, and X is a halide anion. More recently, since 2012, the structure $AMX_3$ (3D perovskites) has led to the very fast emergence of a new class of solar cells, as after a few years of research effort it has led to a conversion efficiency above 23%.[4] However, a number of issues related to hysteretic behaviour, structural, thermal and UV stability, as well as moisture sensitivity have to be solved and a better knowledge of the electronic properties of such materials is obviously a prerequisite for their optimization in different opto-electronic devices.[5,6,7] On the other hand, hybrid perovskite materials have been recently shown to present very efficient radiation hardness together with self-healing properties.[8] The irradiation of solar cell devices with 68 MeV protons allowed to evidence that perovskite absorbers can withstand proton doses up to $10^{12}$ $cm^{-2}$, which exceeds the damage threshold of c-Si by almost 3 orders of magnitude. In the frame of photovoltaic devices, a very good tolerance to electron irradiation (1 MeV, up to $10^{16}/cm^2$) and proton irradiation (50 keV, up to $10^{15}/cm^2$) was also shown.[9] Theoretical calculations revealed that the electronic properties of HOP materials are tolerant to defects because the intrinsic point defects and grain boundaries do not generate gap states, but the origin of this puzzling property is still under debate.[1,10] . It might be also very interesting to use defect engineering in order to better exploit the light-emitting properties of $CH_3NH_3PbI_3$. It was shown for instance that efficient amplified spontaneous emission (ASE) and continuous-wave lasing could be obtained through radiative defect inclusions in polycrystalline thin films in the orthorhombic phase below T=160 K. .[11,12] Interestingly, it was also shown that bulk HOP can be self-doped by defects engineering. It was actually predicted that the electronic conductivity of perovskites can be tuned between p-type and n-type by controlling growth conditions. This self-doping effect in ($CH_3NH_3PbI_3$) was then reported and found to be either n or p type by changing the ratio of methylammonium iodide (MAI) and lead iodine ($PbI_2$) which are the two precursors for perovskite formation.[13] It was inferred that the n-doping comes from the point defects with the lower formation energy, i.e. I vacancies while the p-type behavior was attributed to Pb vacancies.

The firstly used material in the high-performing perovskites solar cells is methylammonium lead triiodide ($CH_3NH_3PbI_3$) which has a tetragonal structure at room temperature (now replaced by triple cations perovskite solar cells). This material undergoes two structural phase transitions as a function of temperature : from orthorhombic to tetragonal at 160 K, and from tetragonal to cubic at about 330 K, showing important impact on charge carrier dynamics and photoluminescent properties.[14]

We use here Helium ion irradiation at different fluences to introduce point defects with varying concentration on purpose in $(CH_3NH_3)PbI_3$ processed both in single crystals and thin films. We compare the photoluminescence (PL) at varying temperature to gain understanding in the radiative recombination mechanisms and the role point defects might be playing.

## Experimental Method

At first, for the materials' synthesis one needs to obtain the methylammonium iodide $(CH_3NH_3)I$ (called MAI hereafter) which is further used as a precursor in the synthesis of the MAPI samples. The MAI was synthesized by adding dropwise 10,5 mL of hydriodic acid HI (57 % in water, stabilized, Sigma Aldrich) to 20 mL of methylamine $CH_3NH_2$ (2 M solution in ethanol, Sigma Aldrich) at 0°C for 2 hours. The solvent was then evaporated at 60°C under vacuum using a rotary evaporator. The powder was subsequently washed several times with diethyl ether and dried overnight at 60°C. In order to increase purity, the yellowish MAI powder was finally recrystallized in a mix of ethanol and diethyl ether.

For thin films synthesis, 159 mg of freshly made MAI and 461 mg of lead Iodide $PbI_2$ were dissolved in 1 mL of N,N-dimethylformamide (DMF). Quartz substrates (Neyco) were cleaned with acetone and ethanol, then treated with a 10 %wt solution of KOH in ethanol in an ultrasonic bath for 15 minute each. The slides were then rinsed with distilled water and dried with pressured air. $MAPbI_3$ thin films were obtained by spin coating the precursor solution at 2000 rpm for 15 seconds. The films were then annealed at 90°C for 20 minutes in air, allowing solvent evaporation as well as thin films crystallization. According to SEM characterizations the average grain size is expected to be about or less than 1 μm. Absorption spectroscopy measurements give an approximate thickness of 400 nm for the obtained polycrystalline thin films.

For single crystals synthesis, 636 mg of MAI and 1844 mg of lead Iodide $PbI_2$ were dissolved in 4 mL of DMF (1:1 molar ratio) in a small Teflon capped vial. The vial was then placed in an oil bath at 105°C. Single crystals of $MAPbI_3$ start to appear at the bottom of the vial after a couple of hours. They were recovered, dried and washed with diethyl ether several times. The typical size of the single crystals is about 3 mm.

The Helium ion irradiation was performed using the IRMA implanter at CSNSM laboratory.[15] We used a rather low ion energy of 30 keV and fluences between $10^{14}$ $cm^{-2}$ and $10^{16}$ $cm^{-2}$. The ion beam size is of the order of 1 mm and is rastered over a large area in order to have a homogeneous dose deposition. These conditions could also be accessible in any laboratory using a commercial ion gun. The implantations were performed at room temperature with ion currents below 2 $\mu A.cm^{-2}$.

The photoluminescence was measured in reflection geometry at 45° using an argon laser ($\lambda$=488nm) at a power of 10 mW as an excitation source, and a spectrometer TRIAX320 from Horiba-JY equipped with a R928 photomultiplier (Hamamatsu) detector. For these measurements the samples were glued with silver paste in an optical closed-cycle cryostat from ARS Instruments equipped with two quartz windows. The time-resolved PL was measured using the Fluoromax equipment (Horiba Jobin-Yvon) based on TCSPC technique with an excitation from a nano-LED at 482 nm pulsed at 100 kHz (full time-range 6 µs).

## Results and Discussion

According to the SRIM simulation program, the mean projected range for the Helium ions in MAPI is 275 nm at 30 keV, and the maximum implanted concentration at the highest fluence (1E16 $cm^{-2}$) is about 0,5 at.%.[16] The light penetration depth in these materials is on the same order of magnitude due to strong optical absorption. Photoluminescence spectroscopy is thus well suited to study the near-surface region that is modified with ion irradiation. The depth distribution of displaced atoms was determined using the detailed calculation with full damage cascade in the SRIM program. At the highest ion fluence (1E16 $cm^{-2}$), the proportion of displaced atoms is about 18 at.% at about 200 nm for the iodine atoms which are the most susceptible to be displaced. The proportion of displaced atoms for the other atomic species are respectively 14%, 6% for H, Pb and 2% for both C and N atoms. So we hereby cover a range of displaced atoms between 0,18% at 1E14 $cm^{-2}$ to 18% at 1E16 $cm^{-2}$ for iodine atoms as estimated from the SRIM simulation.

The photoluminescence from MAPI polycrystalline thin films at room temperature is represented in Figure 1 before and after ion irradiation at 1E14, 1E15 and 1E16 $He^+$ $cm^{-2}$. The reference spectrum is centered at 1.606 eV with a Full Width at Half Maximum of 90 meV, evidencing a contribution from several in-gap states below the band gap energy at about 1.65 eV as obtained from absorption spectrum measurements and according to the known value for this material.[1] After ion irradiation we observe an intensity decrease of the overall photoluminescence spectra, however not very drastic as the intensity observed after irradiation with a high fluence of 1E16 $He^+$ $cm^{-2}$ is only decreased by a factor about 20. As a matter of comparison, an irradiation of crystalline silicon with Argon ions with a number of displaced silicon atoms $10^4$ times lower, ie 0.02 at.%, (equivalent to 1E12 $He^+cm^{-2}$ with the MAPI irradiation conditions used here) decreases the overall photoluminescence intensity together with the lifetime of minority carriers by a factor 100. Using ion irradiation in crystalline silicon with 0.2 at.% displaced atom (equivalent to 1E13 $He^+cm^{-2}$ with MAPI), the signal was not measurable, and lower than for a bare Si wafer.[17] This comparison emphasizes the outstanding radiation hardness of light emission properties : methylammonium lead iodide can withstand ion irradiation fluences about a factor $10^5$ higher than crystalline silicon, which is consistent with what was found by other authors.[9]

It is also remarkable that the PL decay curves are drastically affected only after an irradiation at 1E16 $He^{+}cm^{-2}$ as observed in Figure 2. As a matter of comparison, we derived lifetimes from a multi-exponential model using a sum of 3 exponential functions to describe the data which were fitted up to 6 μs. Before irradiation, the reference sample shows 2 short lifetimes at 37±1 ns (17%), 200±8 ns (37%) and a long lifetime 1097±15 ns (46%). After irradiation at the lowest fluences (1E14 and 1E15 $cm^{-2}$), only the short lifetimes are a little bit affected while the longest lifetime is not changed (34±1 ns (16%), 213±6 ns (42%) and 1097 ±15 ns (42%) at 1E14 $cm^{-2}$, and 35±1 ns (16%), 211±9 ns (35%) and 1140 ±17 ns (49%) at 1E15 $cm^{-2}$). It is very interesting to note that only the short time decay constants are affected at moderate defect concentration, as the long lifetime being the governing one for efficient charge collection in the frame of photovoltaic devices. One can associate these short time decays to non-radiative recombinations introduced through ion irradiation as we also observe a lowering of PL intensity as shown in Figure1. One has to reach 1E16 $cm^{-2}$ to observe a clear impact on the lifetimes with a decrease of all 3 time constants to 21±1 ns (18%), 92±7 ns (38%) and 777 ±23 ns (43%), which is further pronounced at 2E16 $cm^{-2}$ (14±1 ns (19%), 72± ns (37%) and 458 ±25 ns (43%)). As was already noticed, one of the key of MAPI success are its excellent opto-electronic properties even in the presence of defects.[1,12,13,18] When comparing polycrystalline thin films to single crystals samples at low temperature, we even observe an enhancement of the optical emission as can be seen from Figure 3. On the same figure we can also compare the PL spectrum at 10K from an irradiated single crystal (2E15 $He^{+}cm^{-2}$) with a reference, non irradiated single crystal. We observe no change in band gap energy after ion irradiation as the spectra all superimpose on the high energy side. However, we can observe some additional intensity on the low-energy side of the spectrum. A more detailed analysis allows to describe the spectra with a decomposition in 3 components that we named BE1, BE2 and BE3 for the 3 different bound exciton processes centered at three distinct energies below the band-gap, respectively 1.603 eV, 1.590 eV and 1.568 eV, as represented in Figure 4. The reference single crystal spectrum decomposition at T=10K is represented in Figure 4(a) and evidences only broad bound exciton recombination features. Indeed the free exciton signature at the energy gap position has been observed only recently with high quality single crystals, and is generally not present in PL measurements from single crystals which are rather dominated by bound exciton recombination on trap states.[19] One can notice that these components get broader as they are more separated from the gap, indicating a more important energy spread of the associated slightly deeper energy levels. After ion irradiation at 2E15 $He^{+}cm^{-2}$, we can observe that the BE3 component is barely not affected, while one can observe a weight transfer from BE1 to BE2, indicating that BE2 should be related to iodine point defects. Indeed, ion irradiation promotes predominantly iodine point defects, of both vacancy or interstitial types. Following the theoretical calculations of Buin and co-workers, iodine vacancy levels should lie within the valence band. The only possible interpretation for a near-band edge level would then be iodine interstitial levels, most probably associated to the BE2 emission.[18] This increase in BE2 intensity relatively to BE1 leads to an apparent peak position change in the overall spectrum by about 5 meV.

When we compare now the polycrystalline thin film spectrum with the irradiated single crystal in Figure 3, we also observe an important increase in BE2 contribution as can be better seen in Figure 5(a), such that the apparent peak position shift with the reference single crystal is now about 12 meV. The polycrystalline thin films are by nature more defective than the single crystals and it is interesting to notice that this is not a drawback regarding light emission as non-radiative recombination centers are not the dominant ones. In the thin films, the two increasing contributions (BE1 and BE2) are more keen to be related to grain boundaries as well as surface and interface defects and maybe lead and iodine antisites.[18,20]

In Figure 6, we can observe the difference in the PL spectra between the reference polycrystalline thin film and the irradiated films at 1E15 and 1E16 $He^{+}cm^{-2}$. Firstly, at 1E15 $He^{+}cm^{-2}$, the total BE1 and BE2 contribution increases and broadens leading to a relative peak position lowering of about 18 meV (decomposition not shown here), while the BE3 contribution still stays almost the same. Also at the highest fluence of 1E16 $He^{+}cm^{-2}$, as observed in Figure 5(b), this contribution is unaffected indicating a very stable component under ion irradiation. At this fluence, a high relative increase of BE2 is seen, leading to a major spectrum contribution. Also the BE1 contribution is shifted to higher energy at 1.631 eV. The high energy tail of the spectrum overpass the known band-gap energy at this temperature (1.64 eV), related to an important broadening of the BE2 component for this ion irradiation fluence, as evidenced from the spectral decomposition in Figure 5. This broadening is attributed to local disorder and the number of different local halide defect configurations which induces energy level broadening. At this high fluence, we also form clusters of point defects that act as non radiative recombination centers as the integrated intensity of the spectrum is decreased by about 40% as compared to the reference MAPI polycrystalline sample.

We now consider the temperature evolution of the PL spectra between the reference thin film and the irradiated sample at 1E15 $cm^{-2}$ from 10K to 150K as represented in Figure 7. In both cases, the intensity decreases with temperature due to thermal activation of non-radiative recombinations. The complex evolution of the PL spectra was already studied by several authors.[19,21,22,23,24,25] They could for instance observe a difference in the PL spectrum at low temperature when measured with low or high fluence and in general at least 2 components at low temperature that would evolve in a single broad emission feature above 120-130 K.[21] Interestingly, the temperature dependence of the peak position changes behavior in the same range as the structural phase transition at 160 K : from room temperature to about 150 K the energy increases when the temperature is lowered, while it decreases between 150 K and 10

K.[21,25] We also observe for the reference MAPI thin film such a behavior with a progressive change from a high energy peak which dominates the spectra from 10K to 100 K, progressively replaced by a lower energy feature still lowering in energy until 150 K. At temperatures higher than 150 K this component then progressively broaden and increase in energy until it reaches the room temperature spectra shown in Figure 1. After ion irradiation at 1E15 $cm^{-2}$, we observe in Figure 7(b) that the higher energy feature associated to BE1 (see Figure 5) which increases in energy with temperature for the reference MAPI sample has been completely smoothed out from the irradiated spectra. The temperature dependence in Figure 7(b) indicates rather a dominant contribution from BE2 and BE3 that progressively broaden, merge, and finally evolve towards the same kind of single-contribution in the region 100-110 K. Above 100K the spectra both from reference thin film and irradiated sample look very similar, however the intensity after ion irradiation being comparatively larger. Indeed, as we can see from the dependence of the total integrated intensity represented in Figure 8, the intensity from the reference sample has decreased by about 50% from 10K to 150K while it has decreased only by 30% in the case of the irradiated sample. This effect is thus very interesting, however limited to the orthorhombic phase at low temperature, as a steep intensity decrease is observed when approaching the phase transition temperature at 160 K. This step in the temperature dependence of the intensity was observed and discussed by other authors and attributed to electronic confinement in remaining tetragonal inclusions within the orthorhombic phase, as observed with micro-photoluminescence experiments at low temperature.[22,25,26] Very interestingly, we could interpret the intensity amplification at low-temperature as due to a higher number of tetragonal inclusions in the orthorhombic phase after ion irradiation. The irradiation point defects that we created may act as anchoring centers for the tetragonal inclusions. These inclusions, having a lower band-gap than the surrounding orthorhombic phase can act as charge carrier sinks and confine efficiently the excitons.[21,26] This effect may also be used on purpose to exploit more efficiently the radiative defect inclusions in the goal to realize amplified spontaneous emission and continuous-wave lasing that were shown to be associated to the emission from tetragonal phase inclusions .[11,12] For defect engineering, the irradiation should not be too high, as evidenced on figure 8, where the PL intensity is shown to be lowered on the whole temperature range as observed after irradiation at 1E16 $cm^{-2}$. A small PL increase at low temperature (T<100 K) after ion irradiation is also observed with the MAPI single crystals on the same Figure 8, however we observe no intensity step at the structural phase transition. This difference indicates that the film structure influences the phase transition. The dominating emission centers at low temperature originate from surface and interface defects, that we essentially related to the BE2 contribution in the spectral decompositions discussed together with Figure 5.

## Conclusion

Using Helium ion irradiation and photoluminescence spectroscopy at varying temperature, we could show that the methylammonium lead iodide samples, both in single crystals and polycrystalline thin films, present high radiation hardness as they still present light emission up to the highest irradiation fluences used in this work (2E16 $He^{+}cm^{-2}$). We estimated that they can withstand about $10^5$ times the atomic displacements needed to quench the PL in crystalline silicon. The nature of defects giving rise to radiative recombinations and light emission was discussed. The mechanisms are very similar in single crystals and polycrystalline thin films where grain boundaries and intrinsic defects give rise to the same spectra than irradiation defects in single crystals. Ion irradiation defects are also shown to enhance light emission in the low-temperature orthorhombic phase as seen in the dependence of the total PL integrated intensity. This effect, much pronounced in thin films, was related to quantum confinement within tetragonal inclusions that may still be present in the orthorhombic phase at low-temperature (T<160 K). Another effect of the ion irradiation directly observable is the smoothing out of the emission from the less tightly bound exciton (BE1), and a strong reinforcement of the BE2 component that we tentatively attributed to Iodine interstitials when considering theoretical calculations. Finally, this work sheds some light on defects and their influence over opto-electronic properties in methylammonium lead iodide, opening new possibilities for the use of defects engineering as well as doping through ion implantation.

Acknowledgments

This work was supported by the LabEx PALM (ANR-10-LABX-0039-PALM), and IRS MOMENTOM (Université Paris-Saclay).

We acknowledge Cyril Bachelet, Stéphane Renouf and Jérôme Bourçois for the Helium ion irradiation using the IRMA implanter at CSNSM.

Received:
Revised:
Published online:

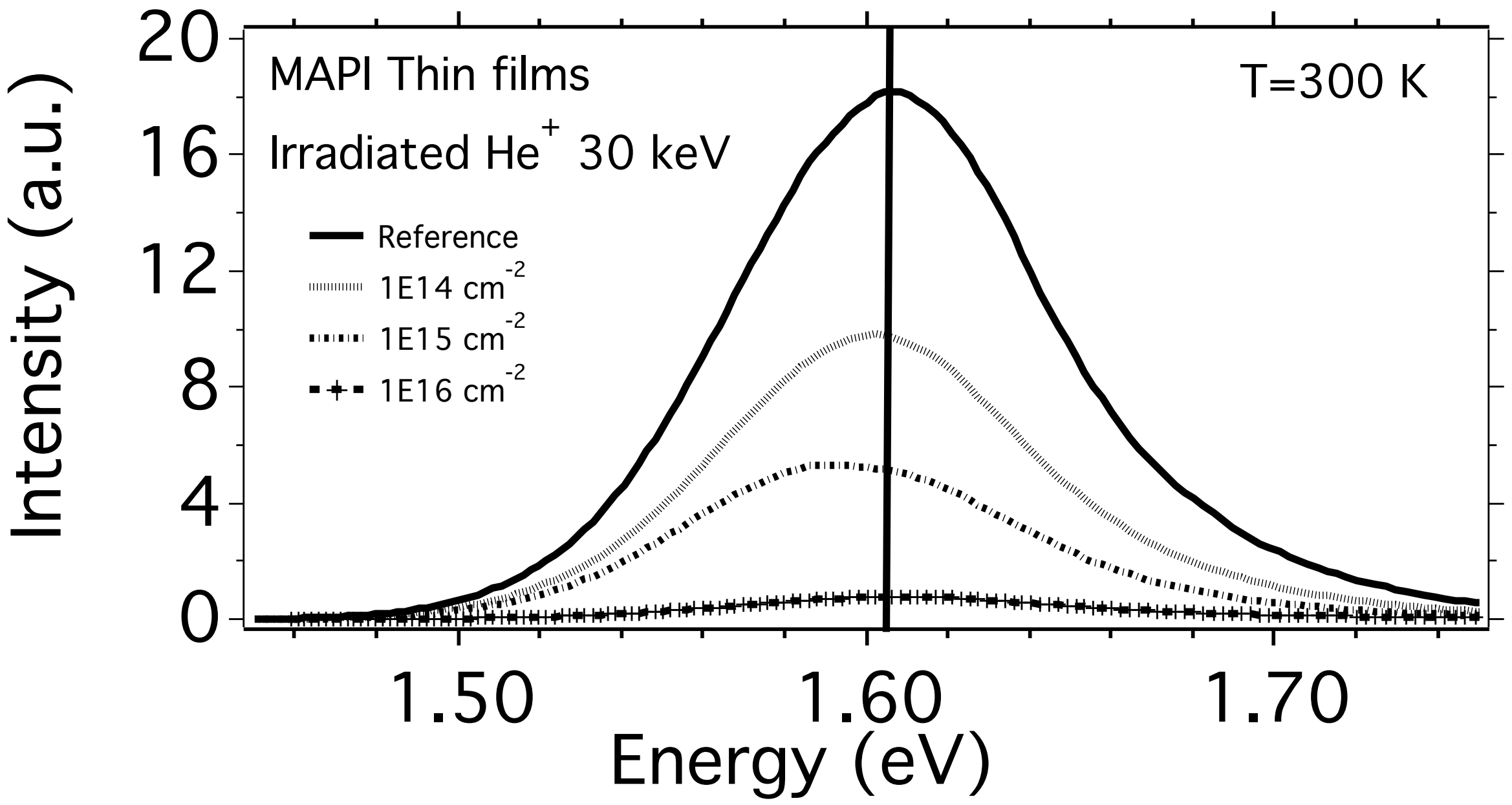

Figure 1 : PL spectra from $(CH_3NH_3)PbI_3$ polycrystalline thin films at room temperature after irradiation with Helium ions at 30 keV and different fluences between $10^{14}$ $cm^{-2}$ and $10^{16}$ $cm^{-2}$.

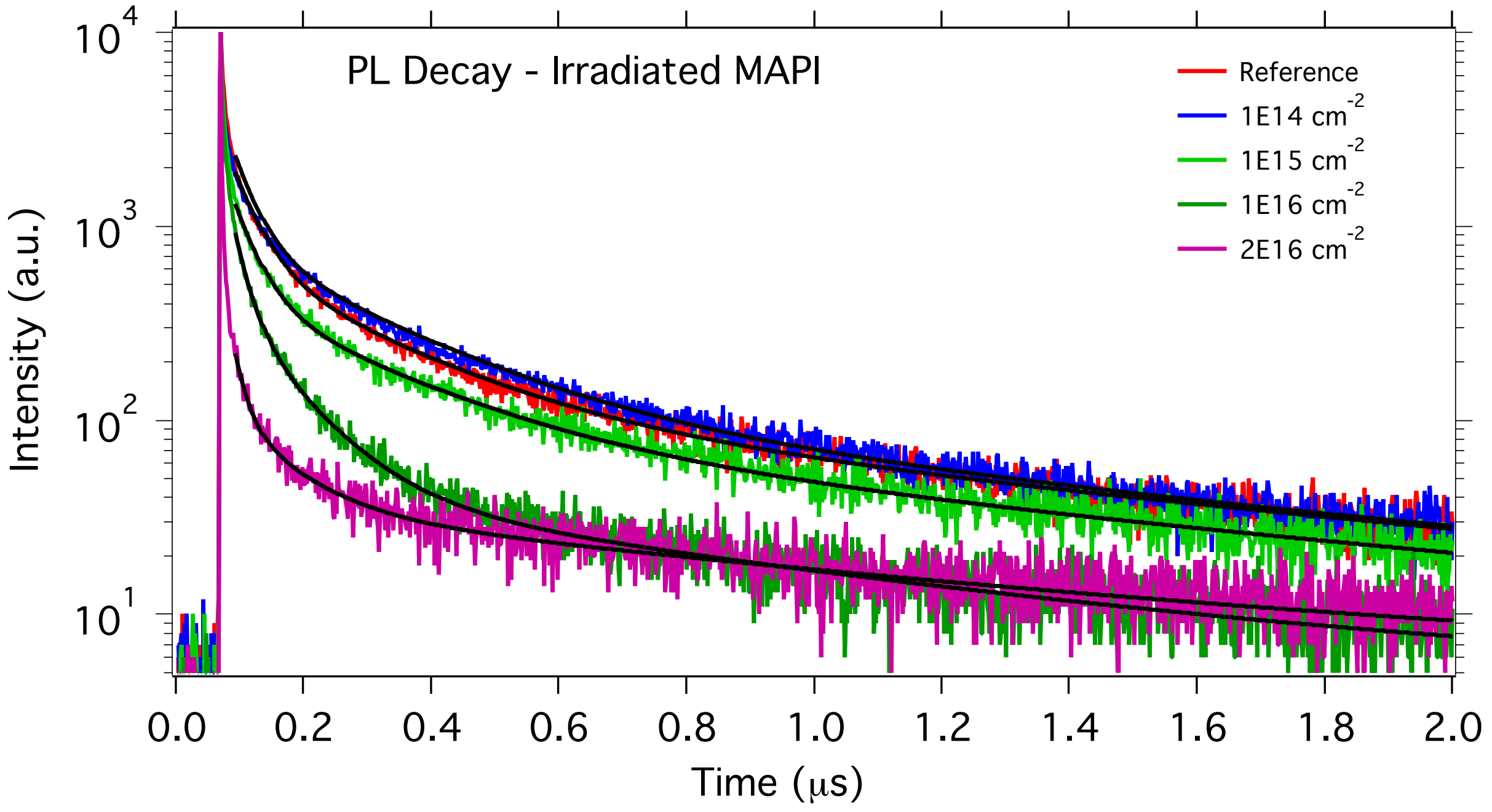

Figure 2 : PL decay from $(CH_3NH_3)PbI_3$ polycrystalline thin films at room temperature after irradiation with Helium ions at 30 keV and different fluences between $10^{14}$ $cm^{-2}$ and $10^{16}$ $cm^{-2}$, measured at 770 nm (1.61 eV). The lines correspond to the best fits using a multi-exponential fit model (3 exponentials).

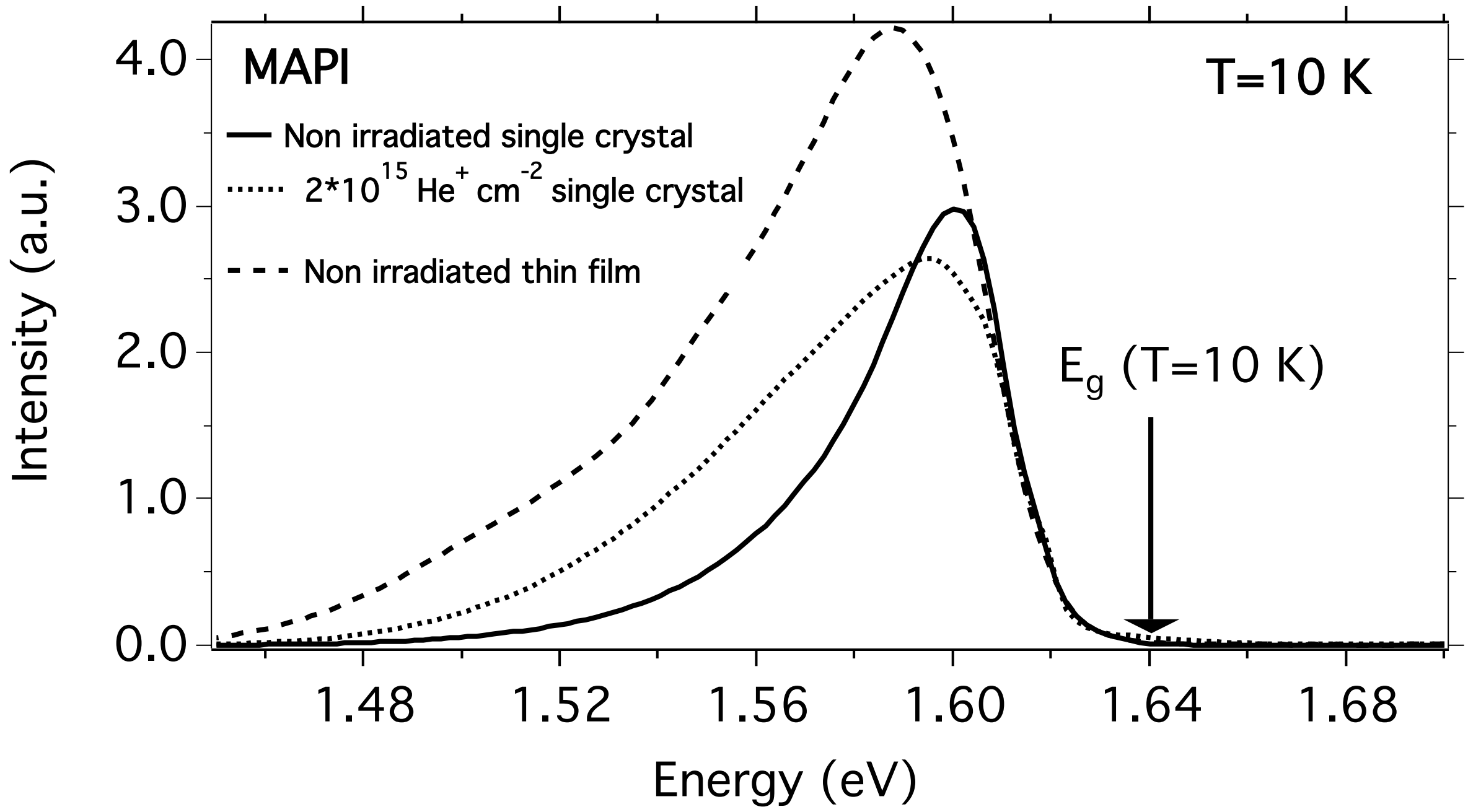

Figure 3: PL spectra from $(CH_3NH_3)PbI_3$ monocrystal and thin film at T=10 K, together with the spectrum from an irradiated monocrystal with Helium ions at 30 keV and 2E15 $cm^{-2}$.

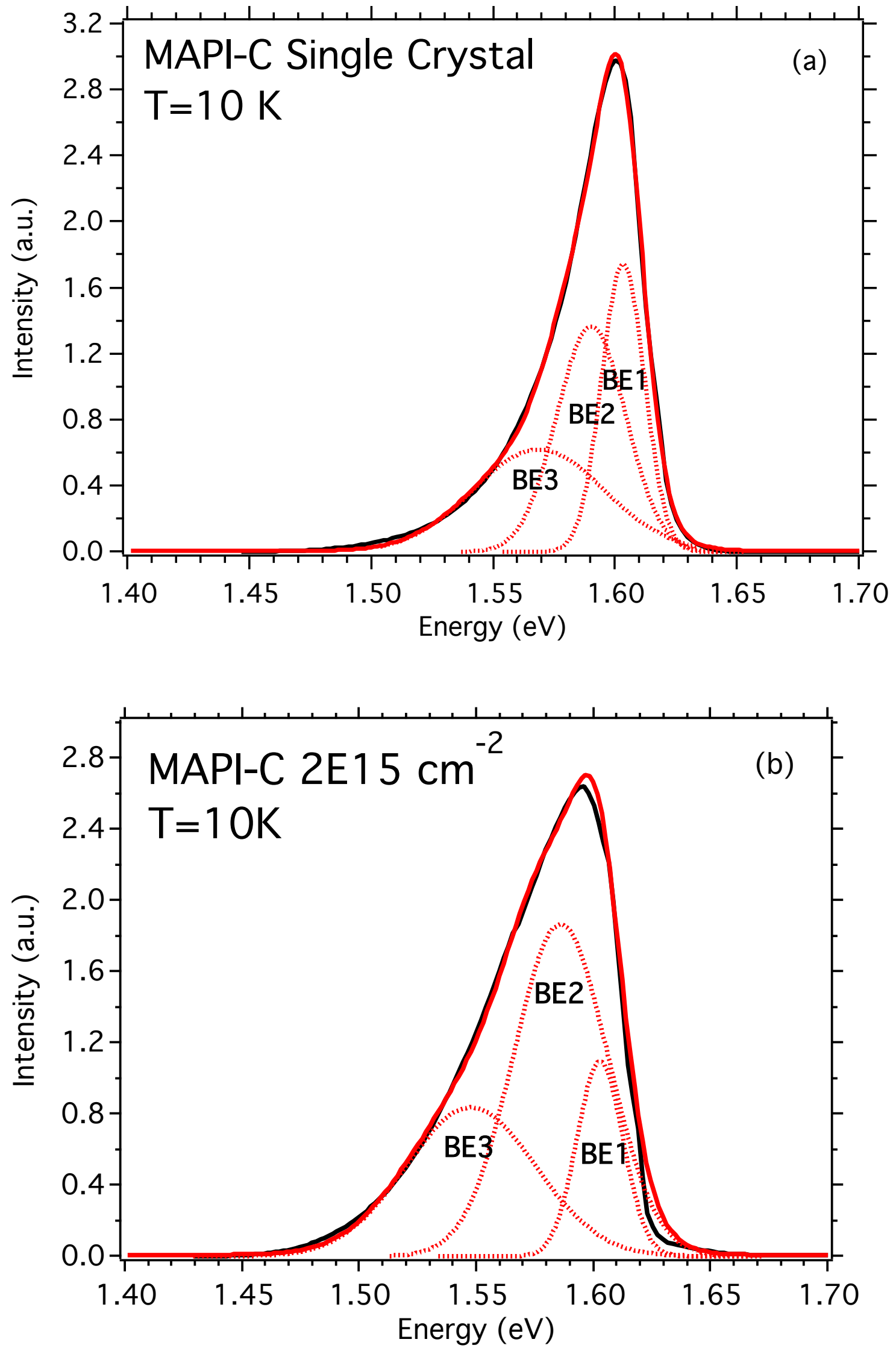

Figure 4: PL spectra from $(CH_3NH_3)PbI_3$ single crystals at T=10 K, before (a) and after irradiation with Helium ions at 30 keV 2E15 $cm^{-2}$ (b). A decomposition of the spectra in three bound exciton components is shown (BE1, BE2 and BE3), together with the fit to the spectra.

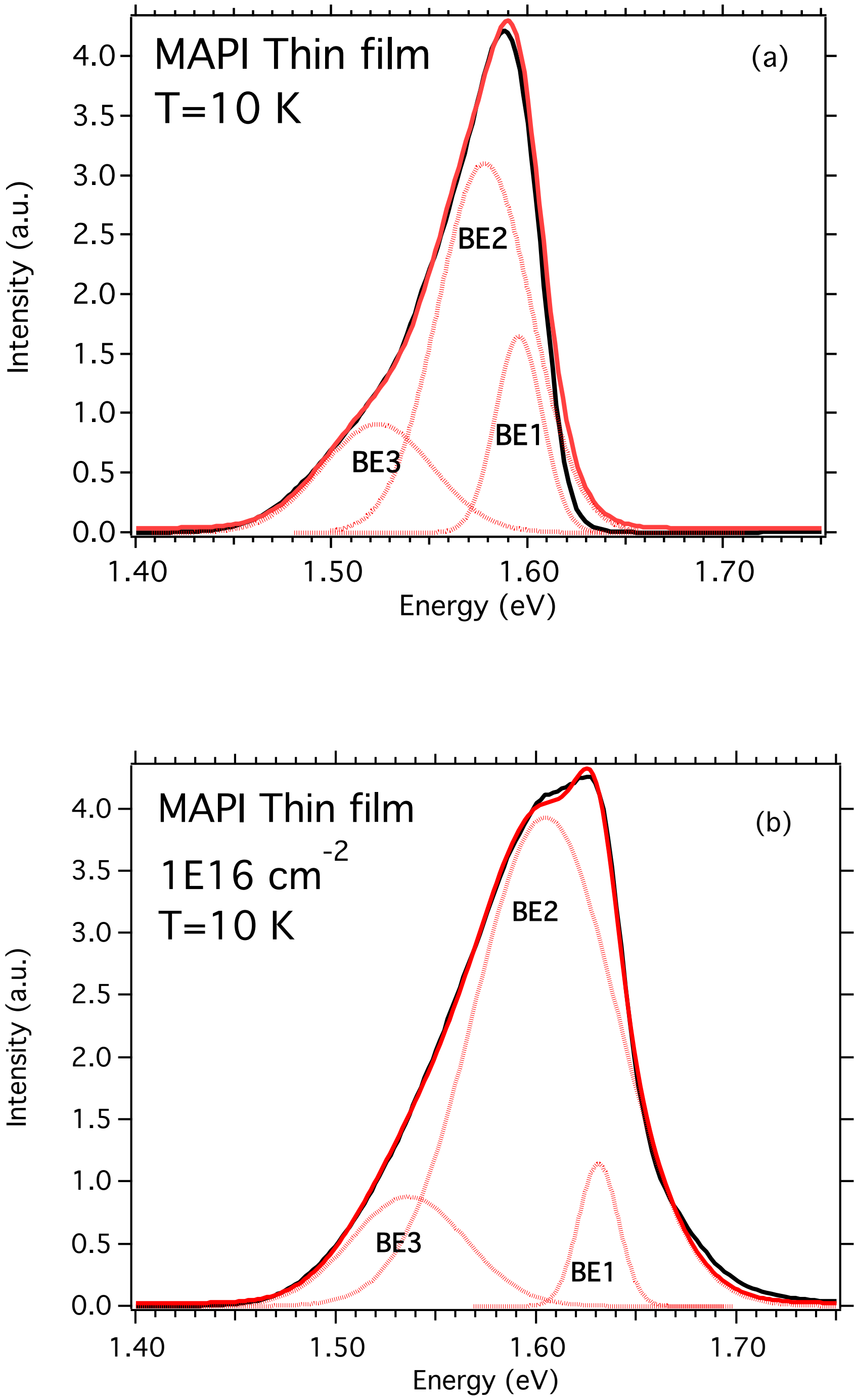

Figure 5: PL spectra from $(CH_3NH_3)PbI_3$ thin films at T=10 K, before (a) and after irradiation with Helium ions at 30 keV 1E16 $cm^{-2}$ (b). A decomposition of the spectra in three bound exciton components is shown (BE1, BE2 and BE3), together with the fit to the spectra.

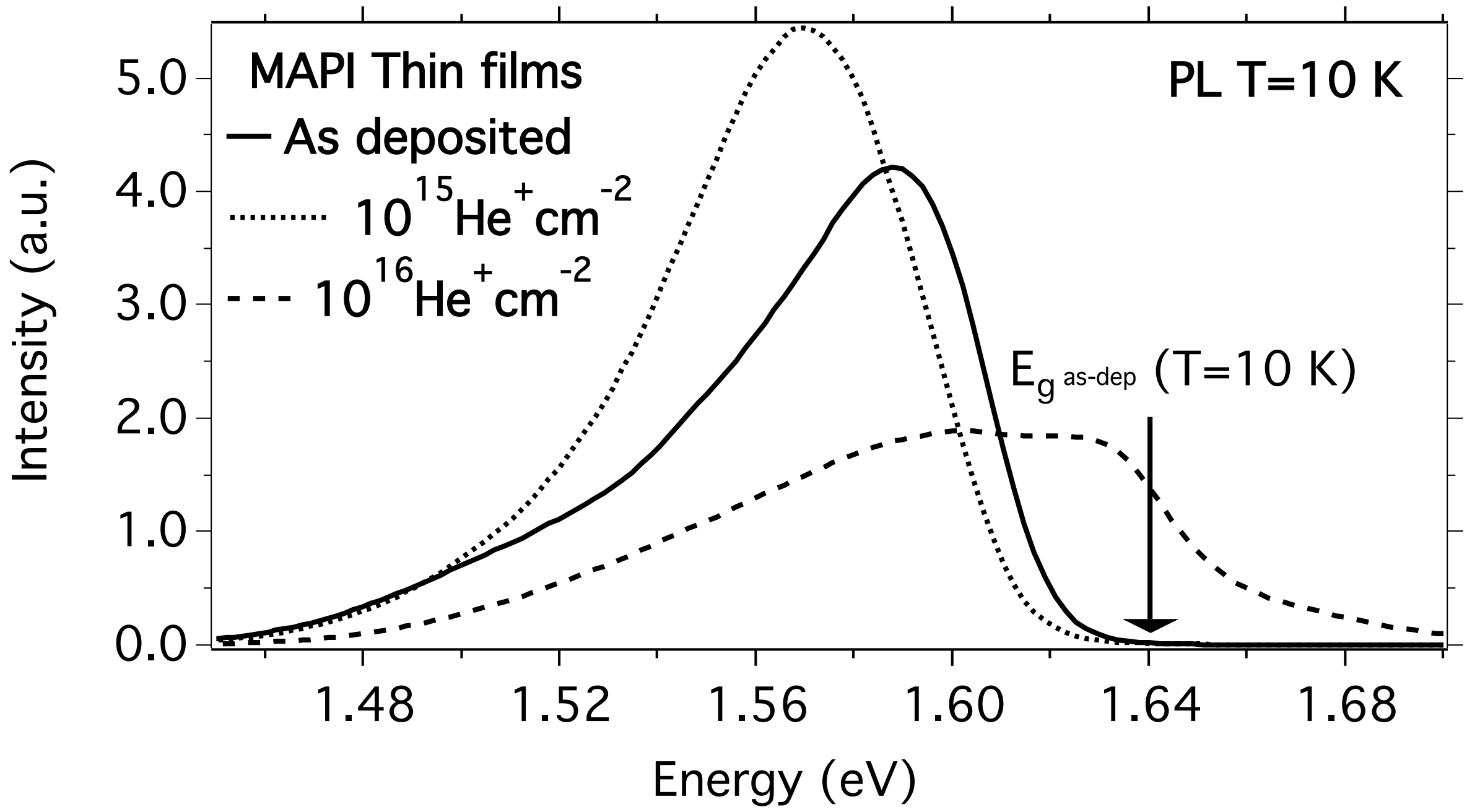

Figure 6: PL spectra from $(CH_3NH_3)PbI_3$ thin films at T=10 K, before and after irradiation with Helium ions at 30 keV 1E15 $cm^{-2}$ and 1E16 $cm^{-2}$.

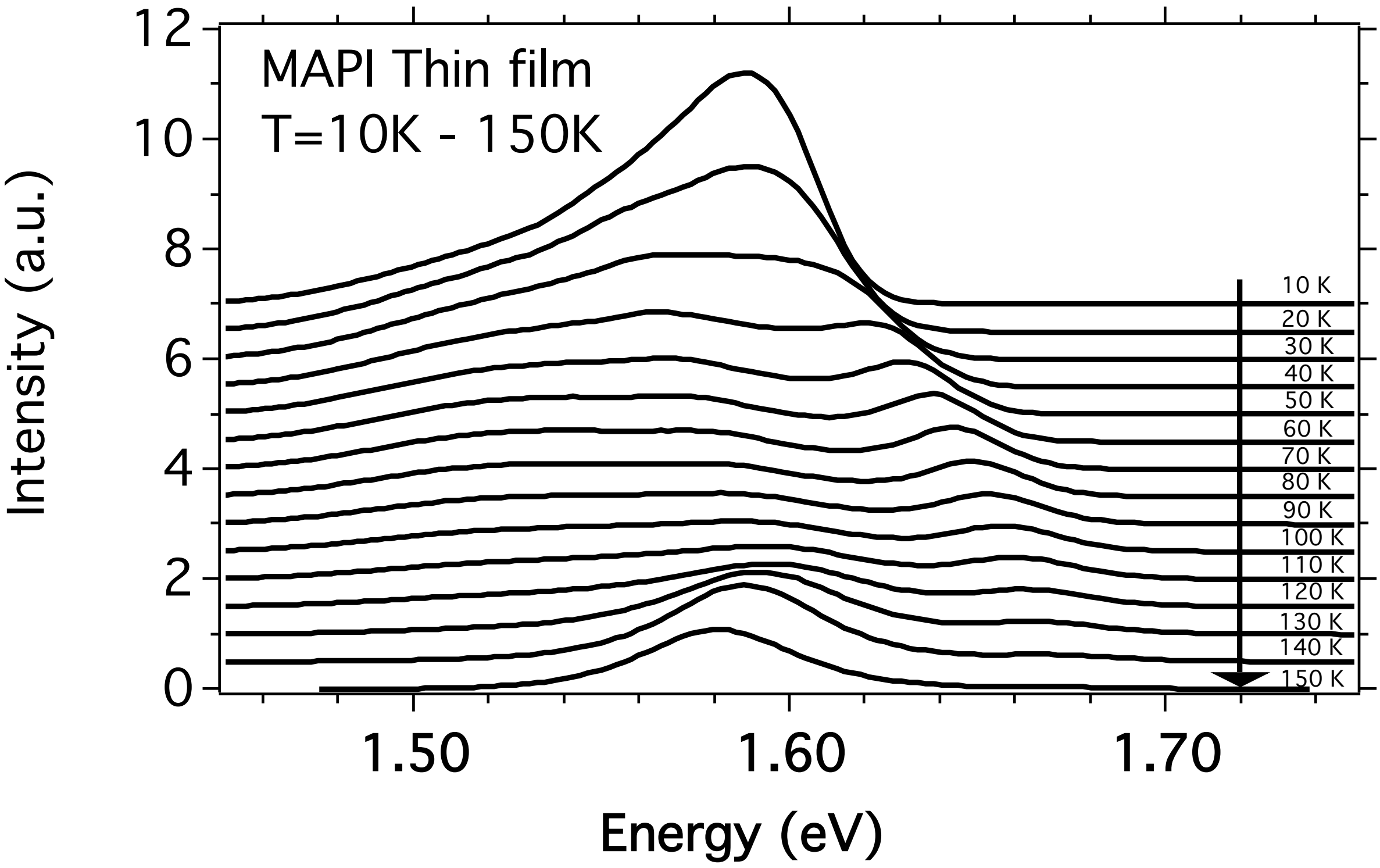

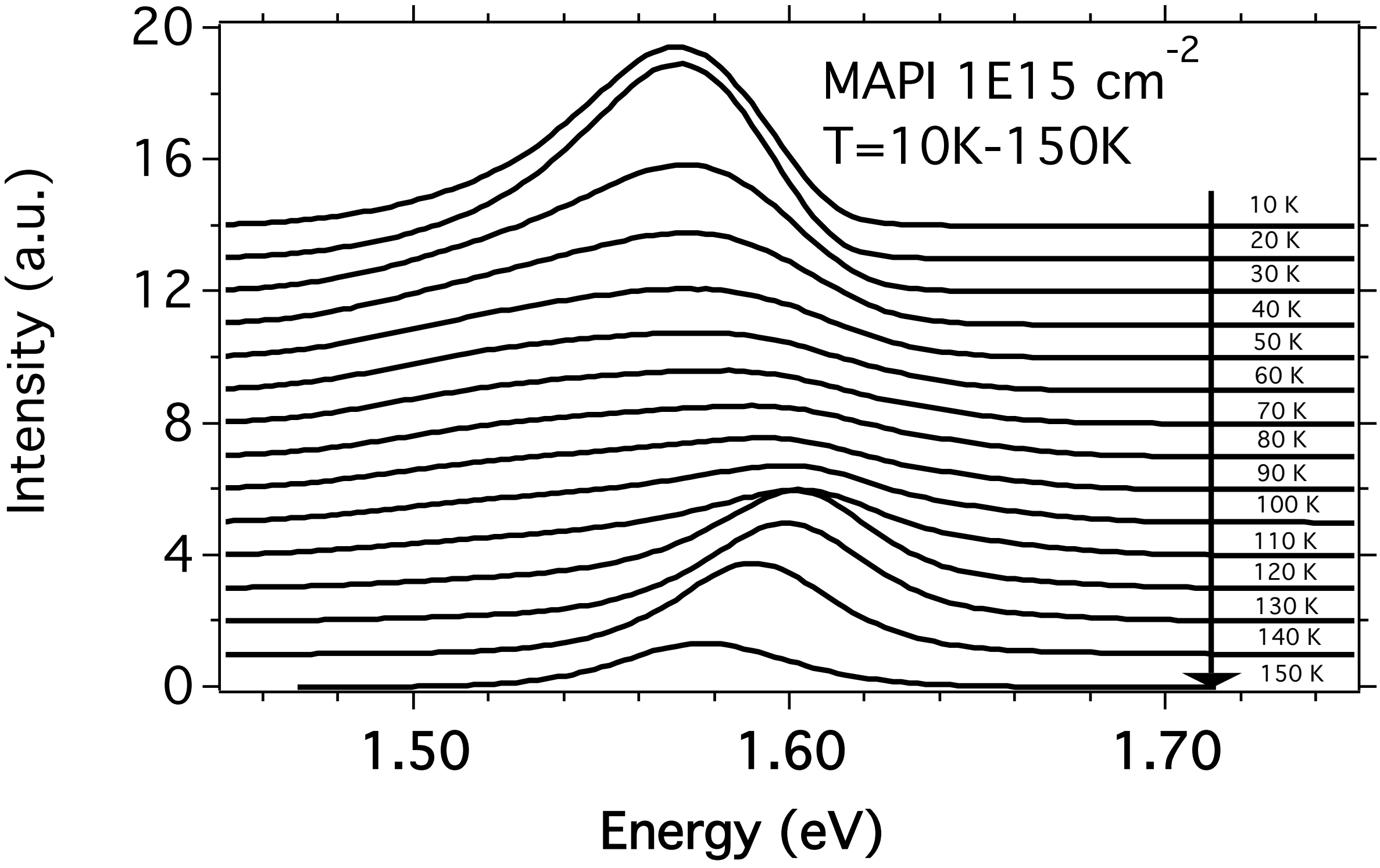

Figure 7: Temperature dependence of the PL spectra from $(CH_3NH_3)PbI_3$ thin films between T=10 K and T=150 K with a 10 K step, before (a) and after irradiation with Helium ions at 30 keV 1E15 $cm^{-2}$ (b).

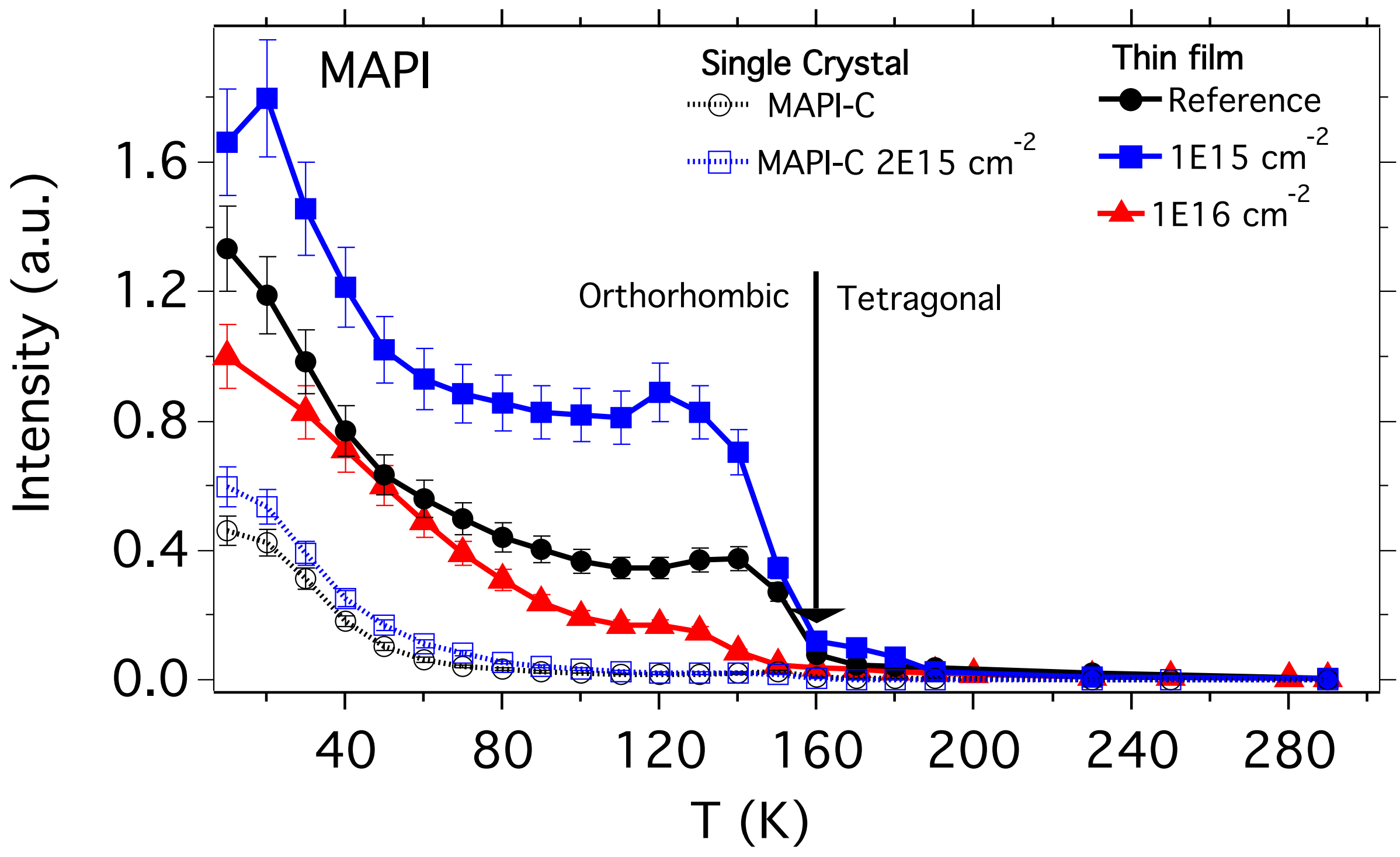

Figure 8: Temperature dependence of the PL integrated intensity from $(CH_3NH_3)PbI_3$ thin films and single crystals, before and after irradiation with Helium ions at 30 keV 1E15 $cm^{-2}$ and 1E16 $cm^{-2}$ for thin films and 2E15 $cm^{-2}$ for the single crystal. The integrated intensity for the reference MAPI thin film was normalized to 1 at T=10 K.